\documentclass[aps,prb,twocolumn,superscriptaddress,showpacs]{revtex4}
\usepackage{ae}
\usepackage[T1]{fontenc}
\usepackage[ansinew]{inputenc}
\usepackage{amsmath}
\usepackage{amssymb}
\usepackage{graphicx}
\usepackage{color}
\usepackage[colorlinks]{hyperref}
\usepackage{epstopdf}

\begin{document}

\date{\today}

\title{Current-induced magnetic superstructures in  exchange-spring devices }

\author{A. M. Kadigrobov}
\affiliation{Department of Physics, University of Gothenburg,
SE-412 96 Gothenburg, Sweden} \affiliation{Theoretische� Physik
III, Ruhr-Universit\"{a}t Bochum, D-44801 Bochum, Germany}
\author{R. I. Shekhter}
\affiliation{Department of Physics, University of Gothenburg,
SE-412 96 Gothenburg, Sweden}
\author{M. Jonson}
\affiliation{Department of Physics, University of Gothenburg,
SE-412 96 Gothenburg, Sweden} \affiliation{SUPA, Institute of
Photonics and Quantum Sciences, Heriot-Watt University, Edinburgh
EH14 4AS, Scotland, UK} \affiliation{Department of Physics,
Division of Quantum Phases and Devices, Konkuk University, Seoul
143-701, Republic of Korea}


\begin{abstract}
We investigate the potential to use a magneto-thermo-electric instability that may be induced in a mesoscopic magnetic multi-layer (F/f/F) to create and control magnetic superstructures. In the studied multilayer two strongly ferromagnetic layers (F) are coupled through a weakly ferromagnetic spacer (f) by an ``exchange spring" with a temperature dependent ``spring constant" that can be varied by Joule heating caused by an electrical dc current. We show that in the current-in-plane (CIP) configuration a distribution of the magnetization, which is homogeneous in the direction of the current flow, is unstable in the presence of an external magnetic field if the length $L$ of the sample in this direction exceeds some critical value $L_c \sim 10$~$\mu$m. This spatial instability results in the spontaneous formation of a moving domain of magnetization directions, the length of which can be controlled by the bias voltage in the limit $L\gg L_c$.
Furthermore, we show that in such a situation the current-voltage characteristics has a plateau with hysteresis loops at its ends and demonstrate that if biased in the plateau region the studied device functions as an exponentially precise current stabilizer.
\end{abstract}

\maketitle

\section{Introduction}
A useful tool for manipulating the local magnetic order in artificially structured materials is provided by the possibility to control the magnetization of  a nanomagnet by injecting an electrical current.  Several such scenarios have been discussed, including the so-called spin torque transfer (STT) technique based on the suggestion by Slonczewski \cite{Slonczewski} and Berger \cite{Berger} to use an injection current of spin-polarized electrons.
The high current densities needed in this case can easily be achieved in electrical point contacts of submicron size, where densities of the order $10^8 \div 10^{10}$~A/cm$^2$ can be reached without significant heating of the material,\cite{Rippard,Yanson} but for larger contacts thermal heating can not be avoided. Instead, Joule heating caused by an (unpolarized) current can be used for thermal manipulation of magnetization. In Refs.~\onlinecite{Jouleheating} and \onlinecite{JAP}, this approach was proposed for varying the strength of the exchange coupling of two strongly ferromagnetic layers (F) separated by a weakly ferromagnetic spacer (f).
In Ref.~\onlinecite{JAP}, the ability of an external magnetic field to change the relative orientation of the magnetization in the outer layers of an F/f/F tri-layer magnetic stack was demonstrated. The result was that by varying the temperature, and hence varying the strength of the exchange-spring coupling through the spacer layer f, the relative orientation of the magnetization of the outer F-layers could be continuously and reversibly changed from being parallel to being antiparallel. Consequently, Joule heating by forcing a dc current through the structure allows an electro-thermal manipulation of the relative magnetization directions.

This kind of dc current-induced manipulation of the magnetization direction was further studied and observed in stacks predicted to have nonlinear  N- and S-shaped current-voltage characteristics (CVC)  in Refs.~\onlinecite{JAP} and \onlinecite{s-shaped}, where temporal oscillations of the magnetization direction, temperature and electric current in the magnetic stack were also investigated.

In this work we explore the possibility to use Joule heating by a current-in-plane (CIP) dc electrical current to control the spatial distribution of the magnetization directions in an exchange-spring layered structure of the type\cite{Davies} sketched in Fig.~\ref{stack}. We will show that, if the voltage bias exceeds a critical value for which the sign of the differential resistance becomes negative, two coupled magnetic domain walls can spontaneously appear along the current flow (see Fig.~\ref{T(x)}) at a distance from each other that can be controlled by the bias voltage.

Domain formation is of course not a new phenomenon. The Gunn effect, \cite{Gunn} well known from the physics of semiconductors, is the name given to the spontaneous formation of (moving) electric domains in a semiconductor biased in a region of negative differential resistance (which requires an N-shaped CVC).  Electric domains in normal metals can also appear and may be due to structural  \cite{Barelko} and magnetic \cite{Landauer,Ross}  transitions, a sharp temperature dependence of the resistance at low temperatures  and magnetic breakdown,\cite{Slutskin,TED,Chiang,Boiko,Abramov1} evaporation \cite{Atrazhev} and melting \cite{Abramov2} (for a review see, e.g., Ref.~\onlinecite{Mints}).
In all these cases, however, the domain sizes are  macroscopically large, typically several cm. In contrast, we will show here that the N- and S-shaped CVC:s of the magnetic exchange-spring structures suggested in Refs.~\onlinecite{Jouleheating,JAP,s-shaped} may give rise to magneto-thermo-electric domains with a characteristic size of the order of 10~$\mu$m.
The spatial distribution of the magnetization in these stacks can vary from one corresponding to a  single magnetic domain wall to a spatially periodic magnetization distribution.

The structure of the paper is as follows. In Sect.~\ref{n-shapedIVC} we briefly discuss some important  features of the temperature dependence of the magnetization orientation in the exchange-coupled stack sketched in Fig.~\ref{stack} and derive the N-shaped CVC  that is the prerequisite for the magneto-thermo-electric instability discussed in Sect.~\ref{MED}. There we show that under certain conditions an instability leads to spatially highly inhomogeneous distributions of the magnetization direction in one of the layers (layer 2 in Fig.~\ref{stack}), of the temperature, and of the electric field inside the magnetic stack, corresponding to the formation of a stable magneto-thermo-electric domain (MTED) structure in the stack. In the concluding Sect.~\ref{conclusion} we summarize the main results of the paper and estimate the values of the relevant parameters that lead to MTED formation.

\section{N-shaped current-voltage characteristics of a magnetic stack under Joule heating \label{n-shapedIVC}}

\begin{figure}
 \centerline{\includegraphics[width=0.7\columnwidth]{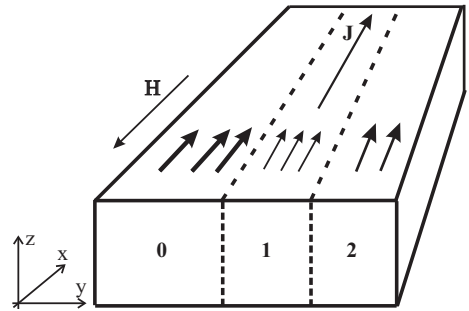}}
 \caption{Orientation of the magnetic moments in the  stack of three ferromagnetic layers
discussed in the text. The magnetic moments in layers 0, 1, and 2 (whose relative magnitudes
are indicated by the thickness of the short arrows) are coupled
by the exchange interaction thus forming an exchange spring tri-layer.
An external magnetic field $H$ is directed antiparallel
to the magnetization in layer 0 and a current $J$
flows in the plane of the layers (along the $x$-axis).}
 \label{stack}
 \end{figure}

 The magnetic stack under consideration has  three ferromagnetic layers
 as shown in Fig.~\ref{stack}. The outer two layers (0 and 2) are strongly
 ferromagnetic and coupled via the exchange interaction through a weakly
 ferromagnetic spacer layer (1).  The Curie temperature $T_c^{(1)}$
 of layer 1 is assumed to be lower than the Curie temperatures $T_c^{(0,2)}$
 of layers 0 and 2. In addition we assume the magnetization direction of layer 0 to be fixed.
A static external magnetic field $H$, directed opposite to the magnetization of layer 0,
is required  to be weak enough that at low temperatures $T$ the magnetization of layer
2 is kept parallel to the magnetization of layer 0 due to the exchange interaction
between them via layer 1. At $H=0$ and $T>T_c^{(1)}$ this tri-layer is similar
to the spin-flip ``free layer" widely used in memory device applications.\cite{Worledge}
The stack is incorporated into an external circuit in such a way that
a current $J$ flows through the cross-section of the layers and
\begin{eqnarray}\label{current}
J= \left[\frac{1}{R\left(\Theta\right)} +\frac{1}{R_0} \right]V\,.
\end{eqnarray}
Here $R\left(\Theta\right)$ and $R_0$ are the magnetoresistance  and
the angle-independent resistance of the stack, $\Theta$ is the
angle between the magnetization directions of layers 0 and 2, and $V$
is the voltage drop across the stack.

\begin{figure}
 \centerline{\includegraphics[width=0.7\columnwidth]{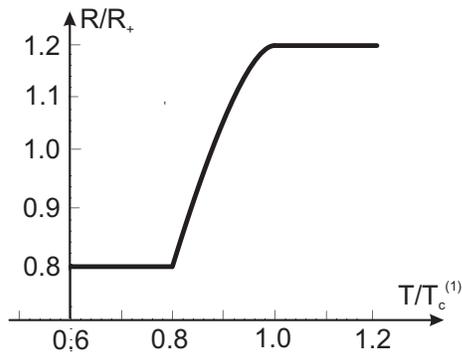}}
  \caption{Normalized temperature  dependence of the magnetic stack resistance, $R=R[\Theta(T)]$, as determined by  the $T$-dependence according to Eq.~(\ref{theta2}) of the angle $\Theta(T)$ between the magnetization directions in layers 0 and 2 of Fig.~\ref{stack}. The calculations were made for $R_-/R_+=0.2$, $D_0=0.2$; $R_{\pm}=R(\pi)\pm R(0)$ and $T_c^{(1)}$ is the Curie temperature of the spacer layer 1 between layers 0 and 2.}
 \label{noflip}
 \end{figure}

In Ref. \onlinecite{JAP} it was shown that a magnetic configuration with parallel orientations of
the magnetization in layers 0, 1 and 2 becomes unstable if the
temperature exceeds some critical temperature $T_c^{\rm(or)} <
T_c^{(1)}$. The magnetization direction in layer 2 smoothly tilts
with an increase of the stack temperature $T$ in the temperature
interval $T_c^{\rm(or)} \leq T \leq T_c^{(1)}$. The dependence of
the  equilibrium tilt angle $\Theta$ between the magnetization
directions of layers 0 and 2 on $T$ and the magnetic field $H$ is
determined by the equation \cite{JAP}
\begin{eqnarray}\label{theta2}
&\Theta&=D(H,T)\sin{\Theta}, \hspace{0.2cm} T < T_c^{(1)} \nonumber \\
&\Theta&=\pm \pi, \hspace{1.9cm} T \geq T_c^{(1)}\,,
\end{eqnarray}
where
\begin{equation}
D(H,T)=\frac{L_1 L_2H M_2(T)}{4 \alpha_1 M_1^2(T)}\approx D_0(H)
\frac{T_c^{(1)}}{T_c^{(1)}-T}.
 \label{D}
\end{equation}
and
\begin{equation}
D_0(H)=\frac{\mu_B H}{k_B T_c^{(1)}}
\Bigl(\frac{L_1}{a}\Bigr)\Bigl(\frac{L_2}{a}\Bigr)
 \label{D_0}
\end{equation}
Here   $L_{1,2}$ and $M_{1,2}(T) $ are the widths  and
the magnetic moments of layers 1 and 2, respectively; $\alpha_1
\sim I_1/ a M_1^2(0)$ is the exchange constant, $I_1$ is the
exchange energy in layer 1, $\mu_B$ is the Bohr magneton, $k_B$ is
Boltzmann's constant,  and $a$ is the lattice spacing. $D(H,T)$ is a
dimensionless parameter that determines how effective the external
magnetic field is to cause the misorientation effect under consideration.
More precisely, it is the ratio between the energy of magnetic layer 2
in the external magnetic field and the energy of the indirect exchange
between layers 0 and 2 (see Fig.~\ref{noflip}).
 At low temperatures the indirect exchange energy
prevails, the parameter $D(H,T) <1$ and Eq.~(\ref{theta2}) has
only one root, $\Theta=0$, thus a parallel orientation of
magnetic moments in layers 0, 1 and 2 of the stack is
thermodynamically stable. However, at temperature $T_c^{\rm(or)} <
T_c^{(1)}$, for which
$$D(T_c^{\rm(or)},H)=1,$$
two new solutions, $\Theta =\pm |\theta_{\rm min}| \neq 0$, appear.
The parallel magnetization corresponding to $\Theta=0$ is now
unstable, and the  direction of the magnetization in region 2
tilts   with an increase of temperature in the interval
$T_c^{\rm(or)} \leq T \leq  T_c^{(1)}$. The critical temperature $T_c^{\rm(or)}$
of this orientational phase transition is obtained from Eq.~(\ref{D}) as\cite{Giovanni}
\begin{eqnarray}
T_c^{\rm(or)}= T_c^{(1)}\left(1- \frac{\delta T}{T_c^{(1)}}
\right), \hspace{0.2cm}
 \frac{\delta T}{T_c^{(1)}}=D_0(H).
 \label{Torient}
\end{eqnarray}

The orientational transition discussed above can be detected by measuring
the temperature dependence of the stack magnetoresistance,\cite{Sebastian}
$R=R[\Theta(T)]$,  plotted for a typical case in Fig.~\ref{T(x)}. This temperature
dependence is caused by the temperature dependence of the misalignment
angle $\Theta=\Theta(T)$, which is implicitly given by Eq.~(\ref{theta2}).

If the stack  is Joule heated  by  a current $J$ its temperature
$T(V)$   is determined by  the heat-balance condition
\begin{equation}
JV=Q(T), \hspace{0.2cm}J  =V/R_{\rm eff}(\Theta),
 \label{heat}
\end{equation}
where
\begin{equation}
R_{\rm eff}(\Theta)=\frac{R(\Theta)R_0}{R(\Theta)+R_0},
 \label{Reff}
\end{equation}
in conjunction with Eq.~(\ref{theta2}), which determines  the temperature dependence
of $\Theta[T(V)]$. Here  $V$ is the voltage drop across the stack,
$Q(T)$ is the heat flux flowing from the stack and $R_{\rm eff}(\Theta)$ is
the total stack magnetoresistance. Here and below we neglect the
explicit dependence of the magnetoresistance on $T$ since we
consider a thin stack  in which  elastic scattering of electrons
is the main mechanism of the stack resistance.
On the other hand, we consider the temperature changes caused
by the Joule heating only in a narrow vicinity of $T_c^{(1)}$, which is
sufficiently lower than both the critical temperatures $T_c^{(0,2)}$ and
the Debye temperature.

Equations~(\ref{heat}) and (\ref{theta2}) define the CVC of the
stack
\begin{equation}
J(V)=\frac{V}{R_{\rm eff}\left[\Theta(V)\right]}, \label{IVC0}
\end{equation}
where $\Theta(V)\equiv \Theta[T(V)]$. Below we will simplify the notation
by dropping the subscript ``eff" and simply write $R(\Theta)$.

From Eq.~(\ref{heat}) it is clear that the differential  conductance of the stack,
$dJ/dV$, is negative if the sample is Joule heated to a temperature at
which $$d \{Q(T)/R[\Theta(T)]\}/dT < 0.$$
Therefore, the main properties of the system under consideration are
determined by the behavior of the function $\chi=Q(T)/R[\Theta(T)]$
plotted for typical parameters in the top panel of Fig.~\ref{chi+X}.
Using Eq.~(\ref{theta2}) one may express the differential conductance
in terms of the dependence of the magnetoresistance $R$  on the
magnetization angle $\Theta$ as\cite{JAP}
\begin{equation}
\frac{dJ_0}{dV}=R \left(\Theta \right) \frac{[R^{-1}\left(\Theta)
(1-{\bar D}\sin{\Theta}/\Theta\right)]'}{[R(\Theta) (1-{\bar
D}\sin{\Theta}/\Theta)]'}\Bigr|_{\Theta=\Theta(V)},
 \label{diffG}
\end{equation}
where
$[\ldots]'$ means the derivative of the bracketed quantity with
respect to $\Theta$, and $${\bar D}=\frac{T}{Q}\frac{d Q}{d T}
D_0\Bigl|_{T=T_c^{(1)}} \approx D_0.$$

\begin{figure}
 \centerline{\includegraphics[width=0.85\columnwidth]{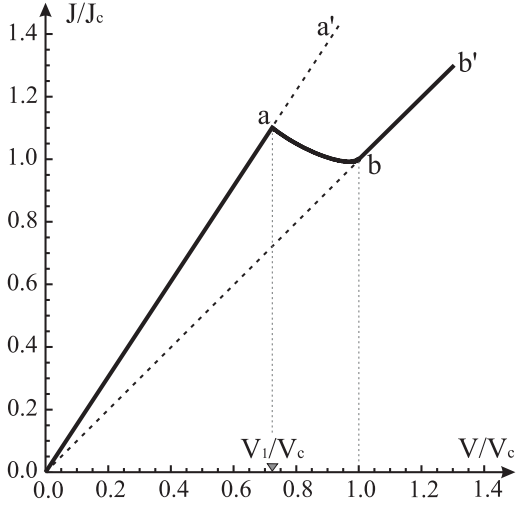}}
\vspace{0 mm} \caption{ Current-voltage characteristics  (CVC) of
the magnetic stack of Fig.~\ref{stack} in which the  magnetization
directions in the layers are homogeneously distributed along the
$x$-direction (along the stack). It was calculated for
$R(\Theta)=R_{+}-R_{-} \cos{\Theta}$, $R_-/R_+ =0.2$, $D_0=0.2$;
$J_c=V_c/R(\pi)$; $V_c =\sqrt{R(\pi)Q(T_c^{(1)})/\Omega_{st}}$.
The branches $0-a$ and $b-b'$ of the CVC correspond to parallel
and antiparallel orientations of the stack magnetization,
respectively (the parts $a-a'$ and $0-b$ are unstable); the branch
$a-b$ corresponds to $0\leq \Theta[T(V)]\leq \pi$.
}
 \label{IVC}
\end{figure}

It follows from Eq.~(\ref{diffG}) the differential conductance
$dJ/dV$ is negative if
\begin{equation}
\frac{d}{d \Theta}\frac{(1-{\bar D}\sin{\Theta}/\Theta)}{R(\Theta)
}<0
 \label{diffGinequality}
\end{equation}
In this case the current-voltage characteristics (CVC) of the stack is N-shaped as shown in Fig.~\ref{IVC}.

Using Eqs.~(\ref{current}) and (\ref{diffG}) and assuming the magnetoresistance to be of the form\cite{Slonczevski1}
$R(\Theta)= R_{+}\left(1 -r \cos{\Theta}\right)$,
where
\begin{equation}
r= \frac{R_{-}}{R_{+}}  >0;\hspace{0.5cm} R_{\pm}=\frac{R(\pi)\pm
R(0)}{2} \label{r}
\end{equation}
one finds that  the
differential conductance $dJ/dV$ is negative if
\begin{equation}
\bar{D} <  \frac{3r\Bigl[ R(0)
-(1-r)^2 R_+ \Bigr]}{(1+2r)R(0)+ (1-r)^2(1-4r)R_+}
 \label{D(CIP}
\end{equation}
It follows that a CVC  with a negative differential
resistance is possible if $R(0) >(1-r)^2 R_+ $.

In the case that the stack has a negative differential
resistance,  nonlinear current and magnetization-direction
oscillations may spontaneously arise if the stack is incorporated
in a voltage biased electrical circuit in series with an inductor.
\cite{JAP,s-shaped} In this paper we show that  another type of
magneto-electrical instability  can arise in such a stack if
the electrical current flows in the plane of the layers
(CIP-configuration):  a homogeneous distribution of
magnetization direction, temperature and electric field along the
spring-type magnetic stack  becomes unstable and a
magneto-thermo-electric domain spontaneous arise in the stack.
Here and below we consider the case that the electrical current
flowing through the sample is lower than the torque critical
current and hence the torque effect is absent.\cite{torquecurrent}

\section{Magneto-electro-thermal instability in a magnetic stack \label{MED}}

In this section we will work in the voltage bias regime, where the resistance
 of the external circuit into which the magnetic stack is incorporated can be neglected in
 comparison with that of the stack. In this case, using the known relation between the electric
 field and the temperature (see, e.g., Ref.~\onlinecite{Landau}) and  taking into account that the
 temperature, $T(x,t)$, being a function of the coordinate $x$ along the stack and
 time $t$, satisfies the continuity equation for the heat flow, one obtains a
set of basic equations for the problem,
 \begin{eqnarray}
&&c_v\frac{\partial T}{\partial t}+j(t)T\frac{d \alpha}{dT}\frac{\partial T}{\partial x}
- \frac{\partial}{\partial x}\Big(\kappa(T) \frac{\partial T}{\partial x}\Big)=-f(T,j)\nonumber\\
&&f(T,j)=Q(T)/\Omega_{st}-j^2(t)\rho [ \theta(T)];\nonumber \\
 &&j(t)\left<\rho \right>=\frac{V}{L}\,,
\label{heateq}
\end{eqnarray}
where the $T$-dependence of $\Theta(T)$ is given by Eq.~(\ref{theta2}). Furthermore, $j(t)=J(t)L/S$ is the current density, which is independent of
$x$ due to the condition of local electrical neutrality, $S$ is the cross-section area of the stack and $L$ is its length, $c_v$ is
the heat capacity per unit volume, $\alpha$ is the proportionality
coefficient between the electric field and the temperature gradient,\cite{Landau} $\kappa$ is the thermal conductivity, $Q(T)$ is the
heat flux flowing from the stack, $\Omega_{st}$ is its volume and
$\rho[\theta]=R_{\rm eff}[\theta]S/L$ is the stack magneto-resistivity
[see Eq.~(\ref{Reff})], the brackets $\langle...\rangle$ that appear in the last part of Eq.~(\ref{heateq}) indicate an average
over $x$ along the whole stack of length  $L$.

The boundary condition needed to solve Eq.~(\ref{heateq}) is the continuity of
the heat flux at both ends of the stack (which is coupled to
an external circuit with a fixed voltage drop $V$ over the stack).
We shall not write down any explicit expression for this condition, since it
will become clear below that the magneto-thermal domain structure
does not depend on the boundary conditions if $L$ is sufficiently large.
Instead, for the sake of simplicity, we use the periodic boundary condition $T(x+L,t)=T(x,t)$.
\begin{figure}
 \centerline{\includegraphics[width=0.85\columnwidth]{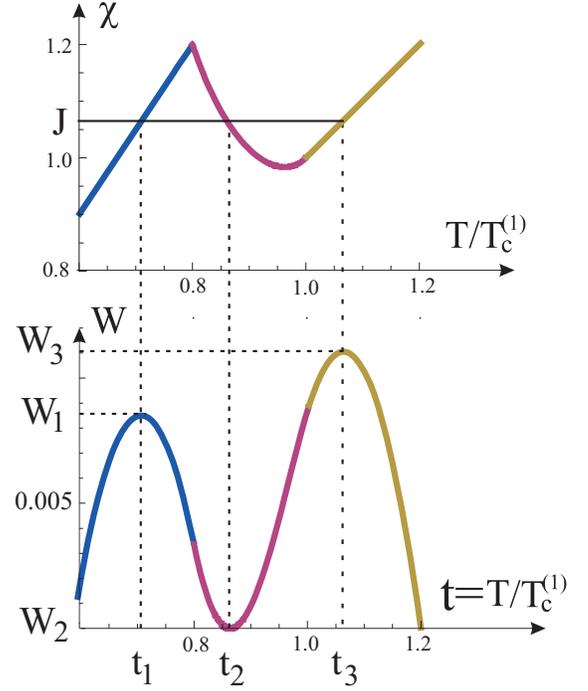}}
\vspace{0 mm} \caption{ Plots of the function $\chi(T)=Q(T)/R[\Theta(T)]$ and the ``potential" energy $W(T,J)$ calculated for
$R(\Theta)=R_{+}-R_{-} \cos{\Theta}$, $R_-/R_+ =0.2$, $D_0=0.2$,
and $J/J_c=1.03$; $J_c=V_c/R(\pi)$.
}
 \label{chi+X}
\end{figure}

The set of equations (\ref{heateq}) always has the steady-state
homogeneous solution
\begin{eqnarray}\label{homogeneous}
&&T=T_0(V), \hspace{0.1cm} \Theta=\Theta_0(V) \equiv \Theta[T_0(V)],\nonumber \\
&&j_0=\rho^{-1}[\Theta (T_0)]V/L,\hspace{0.1cm}f(T_0,j_0)=0.
\end{eqnarray}
 The last equation in (\ref{homogeneous}) is identical to the energy balance condition
 (\ref{heat}) which, together with Eq.~(\ref{theta2}), determines the N-shaped CVC shown
 in Fig.~\ref{IVC}.
\begin{figure}
 \centerline{\includegraphics[width=0.85\columnwidth]{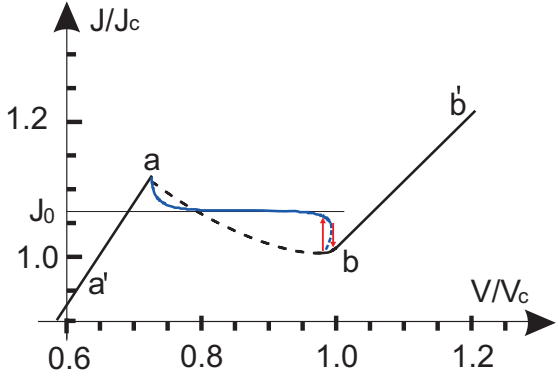}}
\vspace{0 mm} \caption{Current-voltage characteristics (CVC) of
the magnetic stack of Fig.~\ref{stack} for $L\gg L_c$, i.e. when
the stack contains a magneto-thermo-electric domain. The CVC was
calculated for $R(\Theta)=R_{+}-R_{-} \cos{\Theta}$, $R_-/R_+
=0.2$, $D_0=0.2$; $J_c=V_c/R(\pi)$; $V_c
=\sqrt{R(\pi)Q(T_c^{(1)})/\Omega_{st}}$. The branches  $a�-a$
and $b-b'$ of the IVC correspond to parallel and antiparallel
orientations of the stack magnetization, respectively, while the
solid branch $a-b$ corresponds to a stable magneto-thermo-electric
domain (MTED) spontaneously formed  inside the stack. The dashed
lines indicate sections where the CVC is unstable. The red
vertical arrows indicate the hysteresis loop in the CVC caused by
the disappearance and emergence of the  MTED as the bias voltage
is changed. The thin horizontal line shows the stabilization
current $J_{0}$.
}
 \label{IVC+MTED}
\end{figure}

As shown in Appendix~\ref{instability}, if the differential conductance
$dJ/dV$ is negative the uniform magnetization along the stack is
stable only if the stack length $L$ is shorter than some critical length $L_c$, where
\begin{eqnarray}\label{Lcritical}
L_{c}=\sqrt{\frac{2\pi \kappa
\overline{D} T_c^{(1)}}{j^2_c\rho(\pi)}\frac{\rho^{-1}(\Theta_0)
(\sin\Theta/\Theta)^{'}}{\Big[\rho^{-1}(\Theta)\left(1-\overline{D}\sin\Theta/
\Theta\right)\Big]^{'}}}\,
\end{eqnarray}
and the derivatives with respect to $\Theta$ are evaluated for
$\Theta= \Theta_0$. However, when $L> L_c$ the uniform
distribution of the stack magnetization becomes unstable against
fluctuations comprising an arbitrary sum of harmonics $A_n
\exp(i2\pi n x/L)$ ($n=\pm 1,\pm 2...$) with $|n| < L/L_c$.
A fluctuation with $|n|>L/L_c$ or $|n=0|$ (uniform fluctuation), on the other hand,
does not destroy the stability of the homogeneous solution
(\ref{homogeneous}) [see Eq.~(\ref{zeroharmonics})]. We note here
that the characteristic value of the critical length can be rather
short. Using Eqs.~(\ref{Lcritical}) and (\ref{diffG}) and the
Lorentz ratio $\kappa/\sigma =\pi^2k_B^2T/3 e$, where $\sigma=1/\rho$
and $e$ is the electron charge, one finds that
\begin{eqnarray}\label{Lcestimations}
L_c \sim \sqrt{\frac{\kappa }{r}\frac{T_c^{(1)}}{ \rho (0)
j^2_c}}\sim \frac{\pi }{\sqrt{r}}\frac{k_B T_c^{(1)}}{e \rho(0)
j_c} \sim 10\,\mu{\rm m}
\end{eqnarray}
for a realistic experimental situation: $r\sim 0.1\div 0.3$,
$T_c^{(1)}\sim$ 100 K, $\rho (0)\sim 10\,\mu\Omega$cm, $j \sim
10^6\div 10^7$ A/cm$^2$.

Therefore, in the range of parameters $L>L_c$  homogeneous
distributions of the magnetization direction, $\Theta$,
temperature
\begin{eqnarray}\label{TandTheta}
T=T_c^{(1)}\left(1-D_0 \frac{\sin \Theta}{\Theta}\right)
\end{eqnarray}
[see Eq.~(\ref{theta2}) and Eq.~(\ref{D})], and the electric field,

\begin{eqnarray}\label{Electricfield}
{\cal E}=\rho[\Theta(T)]j
\end{eqnarray}
 along the system are unstable, and a
magneto-thermo-electric domain, moving with a constant velocity
$s$ may  spontaneous arise inside the magnetic stack:
\begin{eqnarray}\label{MTED}
&&\Theta(x,t)=\Theta_d(x-st) \equiv \Theta[\nu(x-st,j_d)],\nonumber \\
&&T(x,t)=\nu(x-st, j_d),\hspace{0.1cm} j_d(V)=\frac{V}{L}\Big<
\rho[\Theta_d ]\Big>^{-1}
\end{eqnarray}
where the definition of the brackets $\langle...\rangle$ is the same as in
Eq.~(\ref{heateq}), and $\nu(x)$ satisfies the equation of motion of a
fictitious particle of
``mass" $\kappa \approx \kappa(T_c^{(1)}) $ governed  by a potential
force $f(\nu,j)$ and a friction force proportional to $d\nu/dx$:
 \begin{eqnarray}
\kappa \frac{d^2\nu}{dx^2} +\Big(c_v s -j_d \nu\frac{d \alpha }{d
\nu}\Big)\frac{d \nu}{dx}=f(\nu,j_d) \label{MEDEQ}
\end{eqnarray}
(here $x$ is the ``time" and $\nu$ is the ``coordinate" of the
particle).

 The velocity $s$  of the domain is found from the condition that
 the total change of energy $E(x)$ (see Eq.~(\ref{energychange}))
 in the ``period"~$L$ vanishes:
\begin{eqnarray}\label{velocity} s=j\left\langle \nu \frac{d \alpha}{d
\nu} \left(\frac{d \nu}{dx}\right)^2\right\rangle/ \left\langle c_v(\nu) (\frac{d\nu}{d
x})^2\right\rangle
\end{eqnarray}
Hence, the velocity of a magneto-thermo-electric domain is
$$ s \sim j \alpha /c_v \sim j\frac{1}{c_v T_c^{(1)}}\frac{k_B T_c^{(1)}}{e} \frac{k_B T_c^{(1)}}{\varepsilon_F}\sim
10\,\,{\rm cm/s}$$
for $j\sim 10^6\div10^7$~A/cm$^2$, $c_v \sim 1$~J/K cm$^3$, $T_c^{(1)}\sim$ 100~K.

In the same manner as for Gunn domains in semiconductors \cite{Gunn} and electric domains in superconductors and  normal metals \cite{Mints}, the motion of the domain can be stopped by  inhomogeneities inside the stack  or at its ends. However, there is a possibility that the domain can move along the sample, periodically disappearing at one end of the sample and then reappearing at the other, so that the result is temporal
non-linear electrical oscillations\cite{Mints1} with period $\omega \sim L/s$. In the case under consideration these oscillations involve the magnetization direction $\Theta$, temperature $T$ and electrical current $j$. Using the parameter values already introduced one gets $\omega =s/L \sim 1 \div 10$~MHz for a magnetic stack of length $L\sim 1\,\mu$m.

As shown in Appendix \ref{MEDStructure} (see Eq.~\ref{xi}), the
function $\nu(x)$ satisfies the following
equation with a high accuracy:
\begin{eqnarray}\label{conserveenergy}
\frac{1}{2}\kappa\left(\frac{d \nu}{d x}\right)^2+W(\nu, j_d)=E.
\end{eqnarray}
Here the constant $E$ plays the role of  ``particle" energy, which determines the ``period of motion", $\widetilde{L}(E,j)$, of a non-linear oscillator.
Its value is found from the condition that $\widetilde{L}(E,j)$  is equal to the length $L$ of the magnetic stack, that is  the  magneto-thermo-electric domain $\nu(x)$ as a function of $x$   has only one maximum and one minimum along the length of  the magnetic stack. In the case  $\widetilde{L}(E,j)=L/n$ (where $n=2,3,...$), ``multiple"  domains are also possible in principle. However, they are unstable \cite{TED} and are therefore of no interest for us.

\begin{figure}
 \centerline{\includegraphics[width=0.85\columnwidth]{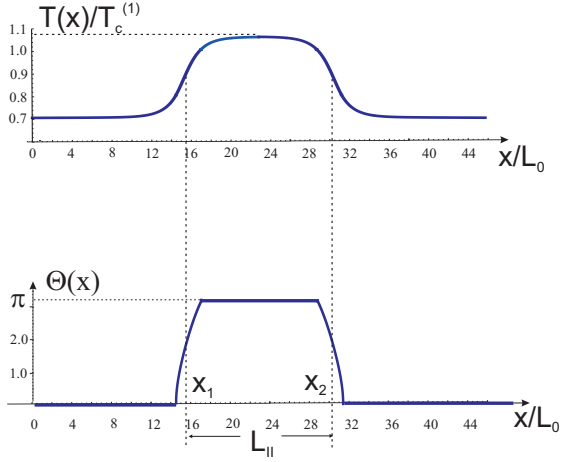}}
\vspace{0 mm} \caption{Coordinate dependence of the temperature,
$T(x)$, and the magnetization-misorientation angle, $\Theta(x)$,
in a magnetic stack containing a magneto-thermo-electric domain,
calculated for $R(\Theta)=R_{+}-R_{-} \cos{\Theta}$, $R_-/R_+
=0.2$, $D_0=0.2$; $J/J_c=1.0265$, $J_c=V_c/R(\pi)$, $V_c
=\sqrt{R(\pi)Q(T_c^{(1)})/\Omega_{st}}$, $L_0
=[j_c^2\rho(\pi)/\kappa T_c^{(1)}]^{(1/2)}\sim L_c$.
}
 \label{T(x)}
\end{figure}

The solution of Eq.~(\ref{conserveenergy}), together with Eq.~(\ref{TandTheta}), defines the spacial distributions of temperature and magnetization direction in a magnetic stack with a magneto-thermo-electric  domain.  Typical examples of such distributions are presented in Fig.~\ref{T(x)}

Using Eq.~(\ref{conserveenergy}) and the last equation in (\ref{MTED}) one finds the set of equations
\begin{eqnarray}\label{period}
\widetilde{L}(E,j)\equiv \sqrt{2 }\int_{\nu_{\rm min}(j)}^{\nu_{\rm max}(j)}\frac{\sqrt{ \kappa}\; d T}{\sqrt{E-W(T,j)}}=L,
\end{eqnarray}
\begin{eqnarray}\label{voltage}
j\sqrt{2 }\int_{\nu_{\rm min}(j)}^{\nu_{\rm max}(j)}\rho\big(\Theta(T)\big)\frac{\sqrt{ \kappa}\; d T}{\sqrt{E-W(T,j)}}=V
\end{eqnarray}
The solutions $E=E_d(V,L)$ and $j=j_d(V,L)$ of these equations are, respectively, the ``energy" of the domain and the current-voltage characteristics of the magnetic stack containing a magneto-thermo-electric domain (the dynamic CVC). In Eqs.~(\ref{period}) and (\ref{voltage}) the limits of integration, $\nu_{\rm min}$ and $\nu_{\rm max}$, i.e., the minimum and maximum temperature of the domain, are obtained as the real roots of the equation $W(\nu, j)=E$.

Let us consider the most pronounced case, for which $L\gg L_{c}$. In this limit, as ``multiple" domains are unstable, one needs  to find a solution of Eq.~(\ref{conserveenergy})  that has a period much larger than $L_{c}$. Since the period $\widetilde{L}(E,j)$ of the non-linear oscillator [see Eq.~(\ref{period})] diverges logarithmically as $E\rightarrow$min$\{W_{1,3}(j)\}$ (where $W_{1,3}(j)$ are the maxima and  minima of the ``potential" energy, see Fig.~\ref{chi+X}),  this condition is satisfied if  the ``energy" $E=E(L,j)$ of the domain  differs from min$\{W_1,W_3\}$ by an exponentially small amount. From here it follows that variations of the current inside a narrow interval  around $j_0$, defined by the relation
$$W_{1}(j_0)=W_{3}(j_0)\,,$$
drastically changes the form of the magneto-thermo-electric domain and hence the current voltage characteristics. Solving the set of equations Eq.~(\ref{period}) and Eq.~(\ref{voltage}) in the interval $|j-j_0|\ll j_0$  (where $W_1 (j)\approx W_3(j) \approx W(j_0)$ ) and taking into account that the maximal, $\nu_{\rm max}$, and minimal, $\nu_{\rm min}$, temperatures of the domain are close to $T_1$ and $T_3$, respectively, one find  the following  implicit form of the dynamic CVC:
\begin{eqnarray}\label{dynamicCVC}
j-j_0 = J_I \exp\Big\{- \frac{\bar{{\cal E}}-\rho(0)j}{r(j_0\rho_{+})}\frac{L}{L_{0}}\Big\}\nonumber  \\
-J_{II} \exp\Big\{\frac{\bar{{\cal E}}-\rho(\pi)j}{r(j_0\rho_{+})}\frac{L}{L_{0}}\Big\}
\end{eqnarray}
Here $\bar{{\cal E}}=V/L$,  $L_0=[f^{'}_T(T_c^{(1)})/\kappa]^{(1/2)}\sim L_c$
and the constants $J_{I,II}$ are of the order of $ j_{0}$ while $f{'}_T=\partial f/\partial T$.

Differentiating both sides of Eq.~(\ref{dynamicCVC}) with respect to $\bar{{\cal E}}$ one finds that the  dynamic CVC  have vertical tangents at the two points $(J_1,\bar{{\cal E}}_1)$ and $(j_2,\bar{{\cal E}}_2)$, where
\begin{eqnarray}\label{j1e1}
&j_1&=j_0\Big(1+\lambda(0)\Big) \nonumber\\
&\bar{{\cal E}}_1&= \rho(0)j_0\Big\{1+ \lambda(0)[1-\ln{\lambda(0)} \Big\}
\end{eqnarray}
and
\begin{eqnarray}\label{j2e2}
&j_2&=j_0\Big(1+\lambda(\pi)\Big) \nonumber\\
&\bar{{\cal E}}_2&= \rho(\pi)j_0\Big\{1-\lambda(\pi)[1-\ln{\lambda(\pi)} \Big\}
\end{eqnarray}
while
\begin{eqnarray}\label{verticaltangents}
\lambda (\Theta)=\frac{L_0}{L}\frac{\rho(\pi)-\rho(0)}{\rho(\Theta)}
\end{eqnarray}
It follows from the above equations that the values of the currents $j_1$ and $j_2$ differ from $j_0$ by an amount $\sim j_0 (L_c/L)$.


One can see from Eq.~(\ref{dynamicCVC}) and Eq.~(\ref{verticaltangents}) that for all values of $\bar{{\cal E}}=V/L$ in the interval $\bar{{\cal E}}_1 <\bar{{\cal E}}<\bar{{\cal E}}_2$ , except for a small region near the ends of the interval, the current $j$ coincides with $j_0$ to an accuracy that is exponential in the parameter $L/L_c\gg 1$.
A change of the bias voltage in this interval does not change the current but it does change the length of the magneto-thermo-electric domain, that is the length of the higher resistive part of it providing the needed voltage drop across the stack at the fixed current value $J=J_0$. Therefore, a magnetic stack with a magneto-thermo-electric domain inside can work as a high-quality current stabilizer.

Near the points  $\bar{{\cal E}}_1 $ and $\bar{{\cal E}}_2$ there is a sharp transition from the nearly horizontal segment of the dynamic CVC to the rising segments of the CVC corresponding to a homogeneous state of the magnetic stack (see Fig.~\ref{IVC+MTED}). As a result, there are hysteresis loops in the current-voltage characteristics of the stack of a large enough length $L$.

The fact, mentioned above, that the current is nearly independent of the bias voltage in the interval $[V_1,V_2]$ (here $V_{1,2}= L {\cal E}_{1,2}$
comes about because in this region the domain structure  can be written
to exponential accuracy (i.e., with an error $\propto\exp(-L/L_c)$)  as
\begin{eqnarray}\label{trapezoid}
T_d(x)=\vartheta(x+x_1)+\vartheta(x_2-x)-T_{\rm max}
\end{eqnarray}
where the function $\vartheta(x)$ is a domain-wall type solution of Eq.~(\ref{MEDEQ}) at $j=j_0$ and $L\rightarrow \infty$, the asymptotic behaviour of which is
\begin{eqnarray}\label{boundary}
\lim_{x\rightarrow -\infty} = \vartheta(x)\rightarrow T_{\rm min} \nonumber \\
\lim_{x\rightarrow \infty} = \vartheta(x)\rightarrow T_{\rm max}
\end{eqnarray}
Here    $x_{1,2}$ are the points of deflections of the curve $d T_d(x)/x$ so that $L_{II} = x_2-x_1$ is approximately the length of the ``hot" section of a trapezoidal  MTED  having the maximal temperature $T_{\rm max}$  and $L_I =L-L$ is the length of its  ``cold" section having  the minimal temperature $T_{\rm min}$ (see Fig.~\ref{T(x)})


\section{Conclusion \label{conclusion}}
We have shown that Joule heating of the magnetic stack sketched in Fig.~\ref{stack} by a current flowing in the plane of  the layers may result in an instability of an initially homogeneous distribution of the magnetization of the stack if the length  $L$ of the stack in the direction of the current flow is longer than some critical length $L_c$. This instability results in the spontaneous appearance of moving domains of magnetization direction in layer 2 of Fig.~\ref{stack}, $\Theta(x-st)$, temperature,  $T(x-st )$ and electric field, ${\cal E}(x-st)$.   For the case $L \gtrsim L_c$ the length of the domain is of the order of $L_c$.

If the length of the stack greatly exceeds the critical length, $L\gg L_c$, the stack is spontaneously divided into two regions, where in one region the magnetization directions in layer 1 and layer 2 of Fig.~\ref{stack} are parallel to each other, while in the other region they are antiparallel. The  length of the region with antiparallel magnetization orientations (that is the length of the domain $L_d$) is controlled by the bias voltage in the  interval $L_c \lesssim L_d \lesssim L$ (see Fig. \ref{T(x)}).
In this case the CVC of a stack containing such a domain has a plateau with hysteresis loops at the ends, as shown in Fig.~\ref{IVC+MTED}. Therefore, the stack can work as a current stabilizer since  the current flowing through it has a fixed value $J_0$ to within an exponentially small error $\propto \exp(-L/L_c)$; a change of bias voltage only results in a change of the domain length to provide the needed voltage drop over the stack.

For a realistic experimental situation the value of the parameter $r$, which through Eq.~(\ref{r}) determines the dependence of the stack resistance on the magnetization-misorientation angle $\Theta$, can be estimated to be of order $0.1\div 0.3$, while the Curie temperature $T_c^{(1)}$ of the spacer layer 1 can be $\sim 100$~K. Using these values and a resistivity of $\rho (\Theta=0)\sim 10$~$\mu\Omega$cm, a current density of $j \sim
10^6\div 10^7$~A/cm$^2$ and a specific heat of $c_v \sim 1$~J/K cm$^3$ one finds that the critical  length is  $L_c\sim 10$~$\mu$m.

\section{Acknowledgements}
 Financial support from the European Commission (FP7-ICT-FET Proj.
No. 225955 STELE), the Swedish VR, and the Korean WCU program
funded by MEST/NFR (R31-2008-000-10057-0) is gratefully
acknowledged.

\appendix

\section{Instability of a magnetization distribution that is spatially homogeneous in the plane of the magnetic layers \label{instability}}

In order to investigate the stability of the spatially homogeneous solution (\ref{homogeneous}) for
temperature, $T=T_0$, current density, $j=j_0$, and magnetization misalignment angle, $\Theta=\Theta_0$, against spatial fluctuations we express these quantities as sums of two terms,
 \begin{eqnarray}\label{corrections}
&T&=T_0(V)+T_1(x,t);\hspace{0.2cm} \Theta=\Theta_0(V)+\Theta_1(x,t)\nonumber\\
&j&=j_0+j_1(t),
\end{eqnarray}
 where $T_1,\Theta_1$ and $j_1$ each is a small correction. Inserting Eq.(\ref{corrections})
 into Eqs.~(\ref{heateq}) and (\ref{theta2}) and  using the Fourier expansion
 \begin{eqnarray}\label{T1}
&T_1(t,x)&=\sum_{n=-\infty}^{+\infty}T_1^{(n)}(t)\exp\{i k_nx\},
\nonumber \\
&\Theta_1(t,x)&=\sum_{n=-\infty}^{+\infty}\Theta_1^{(n)}(t)\exp\{i
k_nx\};\hspace{0.1cm}k_n=\frac{2\pi n}{L}\nonumber
\end{eqnarray}
  one  finds  that $$T_1^{(n)}(t) = D_0T_c^{(1)}|(\sin\Theta_0/\Theta_0)^{'}|\Theta_1^{(n)}(t)$$
  while the  equation for the Fourier harmonics of the angle are
{\setlength\arraycolsep{1pt}
\begin{eqnarray}\label{zeroharmonics}
&&c_v\frac{d \Theta^{(0)}_1}{dt}=-\Big\{\frac{j^2_c\rho(\pi)}{D_0T_c^{(1)}|
(\sin\Theta/\Theta)^{'}|} \nonumber \\
&&\times
\frac{1}{\rho(\Theta)}\Big[\rho(\Theta)
\left(1-\overline{D}\sin\Theta/\Theta\right)\Big]^{'}\Big\}_{\Theta=
\Theta_0}\Theta^{(0)}_1
\end{eqnarray}}
if $n=0$, and {\setlength\arraycolsep{1pt}
\begin{eqnarray}\label{nonzeroharmonics}
&&\frac{c_v}{\kappa}\frac{d \Theta^{(n)}_1}{dt}=- \Big(\frac{2\pi}{L}\Big)^2n^2-\Big\{\frac{j^2_c\rho(\pi)}{\kappa
D_0T_c^{(1)}|(\sin\Theta/\Theta)^{'}|} \nonumber \\
&&\times
\rho(\Theta)\Big[\rho^{-1}(\Theta)\left(1-\overline{D}
\sin\Theta/\Theta\right)\Big]^{'}\Big\}_{\Theta=
\Theta_0}\Theta^{(n)}_1;
\end{eqnarray}}
if  $n \neq 0$.

From this result one sees that if $$[\rho^{-1}(\Theta)\left(1-\overline{D}
\sin\Theta/\Theta\right)]_{\Theta=\Theta_0}^{'}  <0$$  [that is if $dJ/dV<0$, see Eq.~(\ref{diffG})] the uniform magnetization along the stack looses its stability if the stack length exceeds some critical value. Setting  the right-hand side of Eq.~(\ref{nonzeroharmonics}) equal to zero one gets the result (\ref{Lcritical}) for the critical length $L_c$.

\section{Domain structure}\label{MEDStructure}

By multiplying both sides of  Eq.~(\ref{MEDEQ}) by $\partial
\nu/\partial x$ one finds  that the quantity
\begin{eqnarray}\label{totalenergy}
E(x)= \frac{1}{2}\kappa\left(\frac{d \nu}{d x}\right)^2+W(\nu,
j_d)
\end{eqnarray}
plays the role of the total ``energy" of a fictitious particle, the first term giving the
``kinetic energy" and the "potential energy" being defined as
\begin{eqnarray}\label{potentialenergy}
W(T,j)=-\int_{T_2(j)}^T f(T^{'},j)dT^{'}
\end{eqnarray}
(the dependence of $W(T,j)$ on $T$ is shown in Fig.~\ref{chi+X}).
The change of energy with ``time"  $x$ is caused by the action of the
``friction" force:
\begin{eqnarray}\label{energychange}
\frac{d E}{dx}=-\Big(C_v s-\nu (\frac{d\alpha}{d
\nu})^2\Big)\Big(\frac{d\nu}{d x}\Big)^2
\end{eqnarray}

As one sees from Eq.~(\ref{MEDEQ}), the ratio of the ``friction
force" to the ``inertial" term is of the order
\begin{eqnarray}\label{xi}
\xi =j \frac{\alpha L_c}{\kappa }\sim \alpha \sqrt{
\frac{T_c^{(1)}}{r \rho \kappa}} \sim  \frac{k_B T_c^{(1)}}{
\varepsilon_F \sqrt{r}}\,,
\end{eqnarray}
 where $\varepsilon_F$ is the Fermi energy.  From here it follows that the function
 $\nu (x)$ satisfies an ``energy"
conservation law $E(x) =E=const$ (see Eq.~(\ref{totalenergy})) to within
an error  $\sim k_B
T_c^{(1)}/\varepsilon_F \ll 1$, which results in Eq.~(\ref{conserveenergy}).


\begin{thebibliography}{99}
\bibitem{Giovanni} An orientational phase transition in such a system induced by an external
magnetic field was considered in Ref.~\onlinecite{Asti}.
\bibitem{torquecurrent}
Typical current densities needed for the spin transfer torque
effect in point-contact devices are of order $10^8\div 10^9 \rm A/cm^2$ when the current flows perpendicular
to the layers (CPP configuration). \cite{Ralph}
\end{thebibliography}
\end{document}